\documentclass[pre,twocolumn,amsmath,amssymb,nofootinbib,floatfix,superscriptaddress]{revtex4}

\usepackage[normalem]{ulem}
\usepackage{graphicx, bm, tikz, mathrsfs}
\usepackage[breaklinks]{hyperref}

\makeatletter
\def\graphicscale{\twocolumn@sw{0.3}{0.4}}
\def\graphicthreescale{\twocolumn@sw{0.3}{0.4}}

\begin{document}

\title{Effects of quenched disorder in three-dimensional lattice ${\mathbb Z}_2$
  gauge Higgs models}

\author{Claudio Bonati} 
\affiliation{Dipartimento di Fisica dell'Universit\`a di Pisa and INFN Largo Pontecorvo 3,
  I-56127 Pisa Italy}

\author{Ettore Vicari} 
\affiliation{Dipartimento di Fisica dell'Universit\`a di Pisa,
        Largo Pontecorvo 3, I-56127 Pisa, Italy}

\date{\today}

\begin{abstract}
  We study the effects of uncorrelated quenched disorder to the phase
  diagram and continuous transitions of three-dimensional lattice
  ${\mathbb Z}_2$ gauge Higgs models. For this purpose, we consider
  two types of quenched disorder, associated with the sites and
  plaquettes of the cubic lattice.  In both cases, for sufficiently
  weak disorder, the phase diagram remains similar to that of the pure
  system, showing two different phases (one of them being a
  topologically ordered phase), separated by two different continuous
  transition lines.  However, the quenched disorder changes the
  universality classes of the critical behaviors along some of the
  transition lines.  The random-plaquette disorder turns out to be
  relevant along the topological ${\mathbb Z}_2$ gauge transition
  line, so the critical behaviors belong to the different
  random-plaquette $\mathbb{Z}_2$ gauge (RP${\mathbb Z}_2$G)
  universality class with length-scale exponent $\nu=\nu_{\rm
    rp}\approx 0.82$; on the other hand, it turns out to be irrelevant
  along the other Ising$^\times$ transition line (a variant of the
  Ising transitions with a gauge-dependent order parameter), leaving
  unchanged its asymptotic critical behaviors with $\nu=\nu_{\cal
    I}\approx 0.63$.  The random-site disorder leads to a
  substantially different scenario: it destabilizes the Ising$^\times$
  critical behaviors of the pure model, changing them into those of
  the randomly-dilute Ising$^{\times}$ (RDI$^{\times}$) universality
  class with $\nu=\nu_{\rm rdi}\approx 0.68$, while the critical
  behaviors along the other ${\mathbb Z}_2$ gauge topological
  transition line remains stable, with $\nu=\nu_{\cal I}\approx 0.63$.
\end{abstract}

\maketitle

\section{Introduction}
\label{intro}

Statistical systems with quenched disorder are of considerable
theoretical and experimental interest. They model the presence of
impurities whose dynamics is much slower than that of the pure system
variables, thus requiring a quenched average over the free energies at
fixed disorder rather than the standard Gibbs average over the whole
statistical ensemble.  The presence of quenched disorder gives rise to
new phases, such as glassy phases ~\cite{Parisi-nl, Nishimori-book},
as well as novel critical phenomena. In particular, the critical
behaviors in disordered systems can be associated with distinctive
universality classes, which differ from those of the corresponding
pure systems even though they share the same symmetry-breaking
pattern, see, e.g., Refs.~\cite{Ma-book, Cardy-book, Vojta-19}.

Finite-temperature phase transitions and critical behaviors in
disordered systems have been largely studied within lattice spin
models, using various realizations of (typically spatially
uncorrelated) quenched variables, see, e.g., Refs.~\cite{Harris-74,
  HL-74, Khmelnitskii-75, EA-75, GL-76, Nishimori-81, LH-88,PC-99,
  Hartmann-99, PV-00, Betal-00, PV-02, KM-02, KR-03, Granato-04,
  MHT-04, PR-05, KKY-06, Jorg-06, PHP-06, HPPV-07, HPPV-07-rdi, LY-07,
  HPPV-07-mc, HPV-08, HPPV-08, PPV-09, VK-09, FMPTY-09, AV-11,
  BFMMMY-11, CPV-11, Janus-13, BFMS-14, LPP-16, Asp-etal-16, FMMPR-16,
  CP-19, OUK-20, Nishimori-24}. On the other hand, only few studies
have addressed the effects of quenched disorder at the
finite-temperature transitions of statistical systems with gauge
symmetries.  Paradigmatic examples are the finite-temperature
confinement-deconfinement transitions of the three-dimensional (3D)
lattice ${\mathbb Z}_N$ gauge
models~\cite{Wegner-71,Kogut-79,Sachdev-19,BPV-25}, whose topological
nature is related to the fact that they do not present a local order
parameter. Some studies~\cite{DKLP-02,WHP-03,OAIM-04,BV-26}, including
numerical analyses, have been focused on the 3D lattice ${\mathbb
  Z}_2$ gauge models in the presence of spatially uncorrelated
quenched disorder associated with the plaquettes. They show that the
system undergoes continuous confinement-deconfinement topological
transitions even in the presence of a sufficiently weak quenched
disorder. However, as it happens in random-site or random-bond Ising
systems~\cite{Ma-book, Khmelnitskii-75, GL-76, CPPV-04, HPPV-08-2d,
  HPPV-07-mc}, the critical behaviors change, in agreement with the
Harris criterium~\cite{Harris-74,Ma-book,Cardy-book,Vojta-19} (due to the fact
that the specific-heat exponent of the pure system transition is
positive), leading to a distinct random-plaquette ${\mathbb Z}_2$
gauge (RP${\mathbb Z}_2$G) universality class. Indeed, while the
confinement-deconfinement transition of the pure 3D ${\mathbb Z}_2$
gauge system shares the same Ising critical exponent $\nu_{\cal
  I}\approx 0.630$ of the 3D Ising universality class, the critical
exponent of the RP${\mathbb Z}_2$G transition turns out to be
significantly larger~\cite{BV-26}, i.e., $\nu_{\rm rp}=0.82(2)$.

In this paper we extend the study of the effects of quenched disorder
to lattice gauge models coupled to matter fields.  Again, as a
paradigmatic model, we consider a 3D lattice ${\mathbb Z}_2$ gauge
model, adding ${\mathbb Z}_2$ matter variables at the
site of the lattice, which is the so-called 3D lattice ${\mathbb Z}_2$
gauge Higgs model.  The 3D ${\mathbb Z}_2$ gauge Higgs model is
arguably the simplest gauge theory with matter fields, still it shows
a nontrivial phase diagram, characterized by the presence of a
topological phase, and continuous transition lines separating the
normal and topological phases~\cite{Wegner-71, BDI-74, BDI-75, OS-78,
  FS-79, Kogut-79, JSJ-80, HL-91, GGRT-03, Kitaev-03, Nussinov-05,
  CG-08, VDS-09, TKPS-10, GHMS-11, DKOSV-11, CON-11, WDP-12,
  Fradkin-book, Sachdev-19, SSN-21, Grady-21, HSAFG-21, BPV-22-z2g,
  XPK-24, BPV-24, BPV-25}.  This model can also be related to the
quantum two-dimensional toric model in the presence of external {\em
  magnetic} fields, by a quantum-to-classical
mapping~\cite{Wegner-71,Kitaev-03,TKPS-10}, and to a statistical
ensemble of membranes~\cite{HL-91,GHMS-11, SSN-21, SSN-24}.

We study the impact of impurities that can be modelized by
(gauge-invariant) quenched disorders, to understand how they can
change the phase diagram and critical behaviors of the pure ${\mathbb
  Z}_2$ gauge Higgs system. For this purpose we focus on two types of
uncorrelated disorders. One of them is associated with the plaquettes,
like that already considered in the 3D lattice ${\mathbb Z}_2$ gauge
model without matter~\cite{DKLP-02,WHP-03,OAIM-04,BV-26}, while the
other one is associated with the sites where the matter spins are
located, like that considered in the random-site Ising models, see,
e.g., Refs.~\cite{Ma-book,Nishimori-book,PV-02}. Using Monte Carlo
(MC) simulations we show that, for a sufficiently small quenched
disorder, the continuous transition lines of the pure system are still
present, delimiting a topologically ordered phase, although some of
them change the universality class of their critical
behaviors. Moreover, the change of the critical behaviors along the
continuous transition lines crucially depend on the type of quenched
disorder, whether it is of random-plaquette or random-site type.

The paper is organized as follows.  In Sec.~\ref{z2higgs} we summarize
the known features of the lattice ${\mathbb Z}_2$ gauge Higgs model
defined on a 3D cubic lattice, as a necessary starting point to
highlight the effects of the quenched disorder.  Sec.~\ref{withdis}
introduces the models with quenched random-plaquette and random-site
disorder. In Sec.~\ref{numapp} we briefly outline our numerical
approach based on the finite-size scaling (FSS) analysis of the
gauge-invariant energy cumulants computed by MC simulations. In
Sec.~\ref{rplaquette} we discuss how the phase diagram and critical
behaviors change in the presence of random-plaquette quenched
disorder. Sec.~\ref{rdidis} is instead dedicated to the discussion of
the case of random-site disorder.  Finally, in Sec.~\ref{conclu} we
summarize and draw our conclusions.

\section{The three-dimensional ${\mathbb Z}_2$ gauge Higgs model}
\label{z2higgs}

The 3D lattice ${\mathbb Z}_2$ gauge Higgs model is defined by the
Hamiltonian~\cite{Wegner-71, BDI-74, BDI-75, FS-79,Kogut-79}
\begin{eqnarray}
  &&  H = - J \sum_{{\bm x},\mu}
  s_{\bm x} \, \sigma_{{\bm x},\mu} \,
  s_{{\bm x}+\hat{\mu}}
  - K \sum_{{\bm x},\mu>\nu} \Pi_{{\bm
      x},\mu\nu},
\label{HiggsH}\\
&&\Pi_{{\bm x},\mu\nu}=
 \sigma_{{\bm
      x},\mu} \,\sigma_{{\bm x}+\hat{\mu},\nu} \,\sigma_{{\bm
      x}+\hat{\nu},\mu} \,\sigma_{{\bm x},\nu},
\label{plaquette}
\end{eqnarray}
where $s_{\bm x}=\pm 1$ and $\sigma_{{\bm x},\mu}=\pm 1$ are site and
bond variables. The corresponding partition function and free-energy
density are
\begin{eqnarray}
  Z(J,K) &=& \sum_{\{s,\sigma\}} e^{-\beta H(J,K)}, 
    \label{partfuncmodel}\\
  F(J,K) &=& - \frac{1}{V\beta} \ln Z(J,K),
\nonumber
\end{eqnarray}
where $\beta=1/T$ is the inverse temperature, and $V$ is the
volume of the system.  In the following, energies are measured in
units of $T$, which is equivalent to fix $\beta=1$ in
Eq.~\eqref{partfuncmodel}.

An important property of the 3D lattice ${\mathbb Z}_2$ gauge Higgs
model is the existence of a duality mapping~\cite{BDI-75,Savit-80}
between its Hamiltonian parameters. If we define
\begin{eqnarray}
  \left(J^\prime, K^\prime\right)=
  \left( -\frac{1}{2} {\rm ln}\,{\rm
      tanh}\,K, -\frac{1}{2} {\rm ln}\,{\rm tanh}\, J \right),
  \label{dualitymap}
\end{eqnarray}
we obtain a relation between the free energies at different values of
the couplings~\cite{BDI-75}:
\begin{equation}
  F(J^\prime, K^\prime) = F(J,K) - \frac{3}{2}
  \ln[\sinh(2J)\sinh(2K)].
\label{dualityZ}
\end{equation}
This implies that there is a self-dual line,
\begin{equation}
J - J^\prime = J + {1\over 2} {\rm ln}\,{\rm
  tanh}\,K = 0,
\label{selfdual}
\end{equation}
where the duality transformation maps the model into itself, i.e.
$J^\prime = J$ and $K^\prime = K$.  The duality mapping is also
preserved in finite lattices if appropriate boundary conditions are
used, like, e.g., periodic boundary conditions.\footnote{The
$K\to\infty$ limit requires some care: the 3D Ising model is obtained
by setting $\sigma_{{\bm x},\mu}=1$ in Eq.~\eqref{HiggsH}, which
corresponds to $K\to\infty$ in the thermodynamic limit, however at
finite volume nontrivial holonomies can be present even for
$K\to\infty$.  For this reason the dual of the $\mathbb{Z}_2$ gauge
model (corresponding to $J=0$ in Eq.~\eqref{HiggsH}) with periodic
boundary conditions is the 3D Ising model with fluctuating boundary
conditions, see App.  B of Ref.~\cite{BPV-24-coH}.}

\begin{figure}[tbp]
\includegraphics[width=0.95\columnwidth, clip]{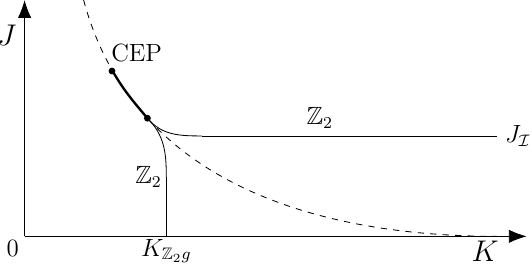}
\caption{Sketch of the $K$-$J$ phase diagram of the 3D lattice
  ${\mathbb Z}_2$ gauge Higgs model (\ref{HiggsH}). The dashed line
  represents the self-dual line. The thick line corresponds to
  first-order transitions on the self-dual line, extending for a
  finite interval.  The two lines labelled ``${\mathbb Z}_2$" are
  related by duality, and correspond to Ising-like continuous
  transitions, whose critical behaviors can be classified as
  Ising$^\times$ with a local gauge-dependent order parameter, which
  is the line ending at $[K=\infty,J_{\cal I} = 0.221654626(5)]$, and
  topological transitions without local order parameters, which is the
  line ending at $[K_{{\mathbb Z}_2g}=0.761413292(11),J=0]$.  The
  first-order transition line gives rise to a critical endpoint
  belonging to the 3D Ising universality class, at $[K_{\rm
      cep}\approx 0.688,J_{\rm cep} \approx 0.258]$.  The three
  transition lines meet at a multicritical point (MCP) on the
  self-dual line, at $[K_\star = 0.7525(1),J_\star = 0.22578(5)]$.  At
  the MCP, the system  develops a multicritical behavior
  characterized by the critical exponents of the XY universality
  class.}
\label{phadia}
\end{figure}

A sketch of the phase diagram is shown in Fig.~\ref{phadia}.  It
presents a topologically ordered phase, delimited by two continuous
Ising-like transition lines that are related by duality.\footnote{In
the context of two-dimensional quantum systems, such a topological
ordered phase is realized in ${\mathbb Z}_2$ spin
liquids~\cite{RC-89,Kivelson-89,RS-91,Wen-91,SF-00,MSF-01}, which is
the phase of matter realized by the toric code~\cite{Kitaev-03}.}
Note that the phases of the model cannot be characterized by their
confinement or deconfinement properties, because the area law of the
Wilson loops is not realized for any $J>0$, due to the screening of
the matter field.  Moreover, the 3D ${\mathbb Z}_2$ gauge Higgs model
presents a further first-order transition line running along the
self-dual line, for a limited range of the Hamiltonian
parameters~\cite{TKPS-10,GGRT-03,JSJ-80}, which does not separate
different phases.

For $K\to\infty$ an Ising transition occurs at~\cite{FXL-18} $J_{\cal
  I} = 0.221654626(5)$.  By duality, a transition occurs at the
corresponding point $J=0$ and $K_{{\mathbb Z}_2g} = -\frac{1}{2} {\rm
  ln}\,{\rm tanh}\,J_{\cal I} = 0.761413292(11)$ (for $J=0$ the model
corresponds to the ${\mathbb Z}_2$ gauge model without matter), which
shares the same Ising critical exponents entering the free-energy
scaling without external magnetic fields, and in particular the
length-scale critical exponent $\nu_{\cal I}$ [see, e.g.,
  Refs.~\cite{PV-02,GZ-98,CPRV-02,KP-17,DBTW-20,Hasenbusch-21,
    Reehorst-22,Chang-etal-25} for accurate numerical estimates of
  Ising critical exponents using different approaches, in
  particular~\cite{Chang-etal-25} $\nu_{\cal I}=0.62997097(12)$
  and~\cite{Reehorst-22} $\omega_{\mathcal{I}}=0.8295(6)$].

Two transition lines, related by the duality transformation (\ref{dualitymap}),
start from the above critical points at $[J=0,K=K_{{\mathbb Z}_2g}]$ and
$[J=J_{\cal I},K=\infty]$~\cite{FS-79, Nussinov-05}.  Their duality relation
implies that they share the same length-scale critical exponent, indeed
$\nu=\nu_{\cal I}$ along both lines.  The two Ising-like transition lines
present however distinct features: the transitions along the line that starts
at $J=0$ are topological, without any local order parameter, while the
transitions along the line that starts at $K = \infty$ are of the
Ising$^{\times}$ type (see the classification proposed in Ref.~\cite{BPV-25}
for the transitions in the presence of gauge symmetries), i.e., they admit an
Ising order parameter that can be identified with the local spin variables
$s_{\bm x}$ once a proper gauge fixing is introduced (such as the stochastic
gauge fixing introduced in
Refs.~\cite{BPV-24-unco, BPV-24-z2gaugeN, BPV-25})\footnote{Other possible order
parameters for these transitions have been proposed and numerically tested,
e.g., in Refs.~\cite{FM-83, FM-86, PK-90, BFZ-98, VBVMRT-22, TRVV-23, XPK-24, SSN-24,
ABP-25}.}.  These Ising-like transition lines intersect on the self-dual line at
a multicritical point (MCP)~\cite{SSN-21,BPV-22-z2g}, located at
$[K_\star=0.7525(1),J_\star= 0.22578(5)]$.

Some numerical studies~\cite{JSJ-80,TKPS-10, GGRT-03} have also provided
evidence of first-order transitions along the self-dual line, in the
relatively small interval, starting from the MCP and ending at
$[K_{\rm ce}\approx 0.688,\,J_{\rm ce}\approx 0.258]$. This endpoint
corresponds to a continuous transition belonging to the Ising
universality class~\cite{SSN-21}.  Since the first-order transition
line is limited to a finite interval along the self-dual line, only
two thermodynamic phases exist \cite{OS-78, FS-79}. For small $J$ and
large $K$ there is a topological phase. The remaining part of the
phase diagram corresponds to a single phase that extends from the
disordered small-$J,K$ region to the whole large-$J$ region. In
particular, no phase transition occurs along the line $K=0$, where the
model becomes trivial (this can be easily seen in the unitary gauge,
in which $s_{\bm x}=1$).  We may only distinguish two different {\em
  regimes}: a Higgs-like regime in the large $J$ and $K$ region, and a
confined regime in the small $J$ and $K$ region (see, e.g.,
Ref.~\cite{FS-79, CG-08}).

At the MCP, where the first-order and the two continuous Ising
transition lines meet, the 3D lattice ${\mathbb Z}_2$ gauge Higgs
model develops multicritical behaviors, which turn out to be
consistent with those predicted by the multicritical XY universality
class~\cite{BPV-25,BPV-22-z2g,XPK-24,ZZ-19}. This scenario can be
associated with the Landau-Ginzburg-Wilson (LGW) $\Phi^4$
theory~\cite{BPV-25,BPV-22-z2g} describing the competition of two real
scalar fields with global ${\mathbb Z}_2\oplus{\mathbb Z}_2$ symmetry,
whose renormalization group (RG) flow has a stable multicritical XY
fixed point where the symmetry gets effectively enlarged to
O(2)~\cite{BPV-22-z2g,BPV-25,CPV-03,NKF-74,FN-74,LF-72}. Therefore,
the two relevant operators driving the multicritical behavior are the
quadratic spin-2 and spin-0 field combinations~\cite{PV-02,CPV-03},
whose RG dimensions are~\cite{Hasenbusch-25-2,Hasenbusch-25} $y_1 =
1.76370(12)$ and $y_2 =1/\nu_{\mathrm{XY}}=1.48872(5)$ (see also
Refs.~\cite{PV-02,GZ-98,CHPV-06,HV-11,KP-17,Hasenbusch-19,CLLPSSV-20}
for other results on the 3D XY universality class). The numerical
results obtained at the MCP of the 3D ${\mathbb Z}_2$ gauge Higgs
model are in good agreement with the above accurate estimates for the
3D XY universality class, indeed Ref.~\cite{SSN-21} reports $y_1 =
1.778(6)$ and $y_2 = 1.495(9)$ (see also \cite{OKGR-23,XPK-24}), and
Ref.~\cite{BPV-22-z2g} reports $y_1 = 1.750(25)$ and $y_2 =
1.495(10)$.\footnote{We mention that the multicritical XY scenario is
not shared by Refs.~\cite{SSN-21, SSN-24, OKGR-23}, which instead
interpret the excellent agreement between numerical results and XY
estimates as a mere coincidence.}

A further interesting characterization of the critical behaviors can
be obtained by the universal features of the relaxational dynamics at
the continuous transitions of the phase diagram. The critical dynamics
is generally characterized by a dynamic exponent $z$, controlling the
power-law divergence of the time correlations at the critical point.
The simplest dynamics is that driven by local purely relaxational
protocols, such as those arising from relaxational Langevin equations
without conservation laws or standard Metropolis upgrading algorithms
in lattice models (model A of the classification reported in
Refs.~\cite{Ma-book,HH-77}). The exponent $z$ generally depends on the
equilibrium universality class of the transition and the type of
dynamics. However, in the presence of gauge symmetries, other features
may turn out to be relevant, as in the case of the relaxational
dynamics of Ising spin and ${\mathbb Z}_2$ gauge
models~\cite{BPV-25-dyn}.

Along the two Ising-like transition lines the purely relaxational
dynamics is characterized by different dynamic exponents, due to the
different nature of the Ising-like transition lines.  Along the
Ising$^\times$ transition line, like any O($N$)$^\times$ vector
transitions~\cite{BPV-25-dyn}, the relaxational dynamic exponent $z$
should be that of the standard Ising universality class, which is
given by~\cite{Hasenbusch-20,FM-06} $z=2.0245(15)$. On the other hand,
along the topological Ising-like transition line, the critical slowing
down developed by the system is expected to be analogous to that of
the 3D lattice ${\mathbb Z}_2$ gauge
model~\cite{BPV-25-dyn,BPV-25-rel,XCMCS-18,BKKLS-90,SPKL-26}, which is
characterized by a significantly larger value of $z$,
i.e.,~\cite{BPV-25-rel} $z=2.610(15)$. In this respect, we note that
since the duality is a nonlocal map, it does not imply a relation
between the local relaxational critical dynamics along the two
Ising-like transition lines~\cite{BPV-25-dyn,BPV-25-rel}.

The relaxational dynamics has been also studied at the MCP,
obtaining~\cite{SSN-21} $z=2.48(10)$.  The above-mentioned
multicritical XY scenario~\cite{BPV-22-z2g}, and the nonlocal nature
of the multicritical modes~\cite{BPV-25}, may lead to the reasonable
hypothesis that the relaxational dynamics at the MCP is analogous to
that of the 3D inverted XY (IXY) universality class, which is that
associated with the continuous transition of the 3D inverted XY gauge
model~\cite{BPV-25} (whose free energy is related by duality to that
of the 3D XY model with Villain action~\cite{DH-81,SSSN-02}).  This
hypothesis is confirmed by the agreement with the estimate $z=2.59(3)$
obtained from numerical analyses of the critical relaxational dynamics
of 3D lattice ${\mathbb Z}_N$ gauge models for $N=6$ and
$N=8$~\cite{BPV-25-zn}, whose finite-temperature topological
transitions belong to the 3D IXY universality
class~\cite{BPV-25}. Note that the IXY exponent associated with the
local relaxational critical dynamics is significantly larger than that
associated with the local relaxational dynamics of the standard XY
universality class, which is given by $z=2.0246(10)$ (obtained in
Ref.~\cite{AEHIKKZ-22} by high-order perturbative computations, see
also Refs.~\cite{FM-06,AV-84,HH-77,HHM-72}).

\section{Lattice ${\mathbb Z}_2$ gauge Higgs model with quenched disorder}
\label{withdis}

In this section we introduce some examples of quenched disorder in the
lattice ${\mathbb Z}_2$ gauge Higgs model, whose phase diagrams will
be discussed in the following sections.

One type of gauge-invariant quenched disorder can be modeled by
attaching a quenched disorder variable $w_{{\bm x},\mu\nu}=\pm 1$ to
the plaquette term, analogously to that already considered for the
lattice ${\mathbb Z}_2$ gauge model without
matter~\cite{DKLP-02,WHP-03, OAIM-04, BV-26}. In this case we may
write the corresponding Hamiltonian as
\begin{eqnarray}
 H = - J \sum_{{\bm x},\mu} s_{\bm x} \, \sigma_{{\bm x},\mu}
 \, s_{{\bm x}+\hat{\mu}} 
- K \sum_{{\bm x},\mu>\nu} w_{{\bm
      x},\mu\nu} \, \Pi_{{\bm x},\mu\nu},
\label{HiggsHrp}
\end{eqnarray}
where $\Pi_{{\bm x},\mu\nu}$ is the plaquette term
(\ref{plaquette}). The quenched disorder variables $w_{{\bm
    x},\mu\nu}$ ($\mu>\nu$) are spatially uncorrelated, and each
disorder configuration is chosen according to the probability
distribution
\begin{equation}
  P_w = \prod_{{\bm x},\mu>\nu} \Bigl[ (1-q)
    \,\delta(w_{{\bm x},\mu\nu} - 1) + q \, \delta(w_{{\bm x},\mu\nu}
    + 1)\Bigr],
\label{probdipla}
\end{equation}
where $0\le q \le 1$. Therefore, $q$ is the probability of getting an
extra minus sign in the Hamiltonian weight of the plaquette.  The
free-energy density $F$ of the system is obtained by averaging the
fixed-disorder free-energy densities $F_{w}$ over the disorder
ensemble $\{w\}$, i.e.,
\begin{eqnarray}
  &&F(\beta,J,K,q) = \sum_{\{w\}} P_w(q,w_{\bm x}) \, F_w(\beta,J,K,w_{\bm
    x}),\label{Fkdis}\\ 
  && F_w(\beta,J,K,w_{\bm x}) = -\frac{1}{\beta V} \ln Z_w(\beta, J, K,w_{\bm x}),
  \nonumber \\
  && Z_w(\beta, J, K, w_{\bm x})=\sum_{\{s,\sigma\}} e^{-\beta H}.  \nonumber
\end{eqnarray}
Again, energies can be measured in units of $T$, which is equivalent
to fix $\beta=1$.

Another interesting possibility is to add quenched disorder in analogy
with the random-site spin models, see, e.g.,
Refs.~\cite{HL-74,Khmelnitskii-75,GL-76,PV-00,HPPV-07-rdi,HPV-07-zrdi}.
For this purpose, we introduce a further random variable $\rho_{\bm
  x}=0,1$ associated with the sites of the lattice, and rewrite the
Hamiltonian as
\begin{eqnarray}
  H = - J \sum_{{\bm x},\mu} \rho_{\bm x} \,\rho_{{\bm
      x}+\hat{\mu}} \, s_{\bm x} \, \sigma_{{\bm x},\mu} \, s_{{\bm
      x}+\hat{\mu}}
  - K \sum_{{\bm x},\mu>\nu} \Pi_{{\bm x},\mu\nu}.\quad
 \label{HiggsHrd}
\end{eqnarray}
The probability distribution of the quenched lattice variables
$\rho_{\bm x}$ is
\begin{equation}
  P_{\rho}(q_s) = \prod_{{\bm x}} \Bigl[ (1-q_s) \,\delta(\rho_{{\bm
        x}} - 1) + q_s \, \delta(\rho_{{\bm x}})\Bigr],
\label{probdisite}
\end{equation}
where $0\le q_s \le 1$ is a global parameter which corresponds to the
impurity concentration when $q_s<1$.  The free-energy density averaged
over disorder is defined analogously to Eq.~(\ref{Fkdis}).

Note that adding a multiplicative random-bond-type quenched disorder
to the bond variables, such as $\sigma_{{\bm x},\mu} = \eta_{{\bm
    x},\mu} \sigma_{{\bm x},\mu}$ with $\eta_{{\bm x},\mu}=\pm 1$ does
not entail any change, because such a disorder can be reabsorbed by a
trivial redefinition of the bond variables, independently of its
distribution.  We also remark that the duality mapping
(\ref{dualityZ}) of the free energies does not hold anymore in the
presence of quenched disorder.

In both random-plaquette and random-site ${\mathbb Z}_2$ gauge Higgs
models, we consider cubic-like systems of size $L$ with periodic
boundary conditions. Of course, one may also consider lattice
${\mathbb Z}_2$ gauge Higgs systems with both types of quenched
disorder. However, we prefer to maintain them distinct, because they
lead to substantially different scenarios.

\section{Finite-size scaling of the energy cumulants}
\label{numapp}

The transitions of the lattice ${\mathbb Z}_2$ gauge Higgs models are
characterized by the absence of local order parameters. Actually, even
the nonlocal order parameter of the lattice ${\mathbb Z}_2$ gauge
models, associated with the area law for the Wilson loops, is not
available for $J>0$, due to the presence of matter fields.  Our
numerical analysis of the phase diagram and critical behaviors of the
3D lattice ${\mathbb Z}_2$ gauge Higgs models in the presence of
quenched disorder, such as those modelized by the
Hamiltonians~(\ref{HiggsHrp}) and (\ref{HiggsHrd}), is based on the
analysis of the finite-size scaling (FSS) of the gauge-invariant
energy cumulants averaged over disorder.

More precisely, we consider the cumulants $B_k$ of
\begin{equation}
  W = - H,
  \label{wdef}
\end{equation}  
which can be defined as
\begin{eqnarray}
B_k(\beta,J,K,q) =\frac{1}{V}\sum_{\{w\}} P_w(q,w_{\bm x}) \,
\left(\frac{\partial}{\partial{\beta}}\right)^k \ln Z_w.
\label{bkdef}
\end{eqnarray}
An analogous definition applies to the model with random dilution of
the site spins, cf. Eq.~(\ref{HiggsHrd}). As already mentioned, after
derivation with respect to $\beta$, we set $\beta=1$, which
corresponds to write $J$ and $K$ in units of the temperature $T$.

One can easily see that $B_1=[\langle W\rangle]/V$ is related to the
averaged energy density $E=[\langle H\rangle]/V$ (they differ for the
sign), where the square brackets $[\phantom{a}]$ indicate the average
over the quenched disorder.  The cumulants for $k>1$ can be related to
the central moments of the energy defined for each disorder
realization by
\begin{equation}
  m_k = \langle \, (W - \langle W\rangle)^k \rangle,
  \label{mkmom}
\end{equation}
where the statistical average $\langle\phantom{a}\rangle$ is taken
over the spin variables at fixed disorder configuration.  The second
and third cumulants are given by
\begin{eqnarray}
B_k=\frac{1}{V} [m_k] \quad{\rm for}\;\;k=2,3.
\label{B23def}  
\end{eqnarray}
In particular, $B_2$ is proportional to the specific heat. The
relation between cumulants and central moments becomes more
complicated for higher cumulants, for example
\begin{equation}
B_4 = \frac{1}{V} [m_4 - 3 m_2^2].
\label{B4def}
\end{equation}

The cumulants $B_k$ are very useful to characterize topological
transitions in which no local gauge-invariant order parameter is
present (see, e.g., Refs.~\cite{SSNHS-03, BPV-20-hcAH, BPV-22-z2g,
  BPV-24-coH, BPV-24, BV-26}), since for fixed $q$ they are expected
to show peculiar FSS behavior~\cite{SSNHS-03, BPV-24-coH}.  For
example, assuming the disorder parameter $q$ and $K$ fixed, they are
expected to scale as
\begin{eqnarray} 
&&B_k \approx L^{k/\nu-3} {\cal B}_k(X) \left[1 + L^{-\omega}
    {\cal B}_{k,\omega}(X)\right] + b_k,
\nonumber\\
&&X = (J-J_c)L^{1/\nu},
\label{Hg3-scaling}
\end{eqnarray}
where the constant $b_k$ represents the analytic
background~\cite{PV-02,BPV-24-coH}, and the $O(L^{-\omega})$
suppressed term, with $\omega>0$, represents the leading scaling
correction, which is generally associated with the leading irrelevant
RG perturbation at the fixed point~\cite{PV-02} (for example
$\omega\approx 0.83$ at the standard 3D Ising transition).  The
scaling functions ${\cal B}_k(X)$ and ${\cal B}_{k,\omega}(X)$ are
universal apart from a multiplicative factor and a normalization of
the argument, however they generally depend on the boundary conditions
adopted.

It is important to stress that the background term $b_k$ in
Eq.~(\ref{Hg3-scaling}) turns out to be subleading with respect to the
scaling term only when $k-3\nu >0$, thus for $k\ge 2$ when the
specific-heat exponent is positive, i.e., $\alpha=2-3\nu>0$.  If
instead $\alpha<0$, higher cumulants are needed to identify the
asymptotic scaling behavior, with the minimum value of $k$ depending
on the critical exponent $\nu$.  Note however that higher cumulants
also become more noisy, so it is typically not convenient to use
values of $k$ much larger than the smallest one which realizes $k-3\nu
>0$. It is also important to remember that, to get an unbiased
estimate the $k$-th cumulant, we need $k$ independent simulations
performed for each noise realization, see, e.g.,
Ref.~\cite{HPPV-07-rdi}. For these reasons, and based on our previous
experience with the $\mathbb{Z}_2$ gauge model \cite{BV-26}, our
simulations were targeted at estimating just the first four energy
cumulants.

\section{Phase diagram in the presence of random-plaquette disorder}
\label{rplaquette}

We now discuss how the phase diagram and critical behaviors of the 3D
lattice ${\mathbb Z}_2$ gauge Higgs model, sketched in
Fig.~\ref{phadia}, changes in the presence of quenched disorder
modeled by the Hamiltonian (\ref{HiggsHrp}).

For this purpose we first focus on the line $J=0$, where we recover
the 3D lattice ${\mathbb Z}_2$ gauge model with a gauge-invariant
quenched disorder associated with the plaquettes, whose Hamiltonian
reads~\cite{DKLP-02}
\begin{eqnarray}
  H_{\rm rp} = - K \sum_{{\bm x},\mu>\nu} w_{{\bm x},\mu\nu} \,
  \Pi_{{\bm x},\mu\nu},
  \label{rpgm}
\end{eqnarray}
with the probability distribution (\ref{probdipla}) for the quenched
variables $w_{{\bm x},\mu\nu}$. Its $q$-$K$ phase diagram, and
critical behaviors at their confinement-deconfinement transitions,
have been studied in Refs.~\cite{DKLP-02, WHP-03, OAIM-04, BV-26}.
This system undergoes continuous topological transitions for
sufficiently small values of the disorder parameter $q$, i.e., $0\le
q\lesssim 0.03$~\cite{DKLP-02,WHP-03,OAIM-04} (such a limit
corresponds to the location of the transition point along the
so-called Nishimori
line~\cite{Nishimori-book,Nishimori-81,WHP-03}).\footnote{For
comparison, we mention that the typical value of the plaquette at the
transition of the $\mathbb{Z}_2$ gauge model is $\langle \Pi_{{\bm
    x}}\rangle\approx 0.95$, see \cite{BPV-25-rel}, thus on average
approximately $5\%$ of the plaquettes have a negative sign at the
transition when no disorder is present.}  Since the specific-heat
exponent of the pure ($q=0$) system is positive (it equals that of the
Ising universality class, $\alpha_{\cal I}\approx 0.11$), according to
the Harris criterium~\cite{Harris-74,Ma-book,Cardy-book,Vojta-19} the quenched
disorder coupled to the plaquette represents a relevant perturbation
that changes the universality class of the critical behavior. Indeed,
a new 3D random-plaquette ${\mathbb Z}_2$ gauge (RP${\mathbb Z}_2$G)
universality class emerges at the continuous transitions for $q>0$,
characterized by the critical exponent~\cite{BV-26} $\nu_{\rm rp} =
0.82(2)$ (obtained from numerical FSS analyses of MC simulations at
$q=0.015$ and $q\approx 0.022$).

\begin{figure}
  \includegraphics[width=0.95\columnwidth]{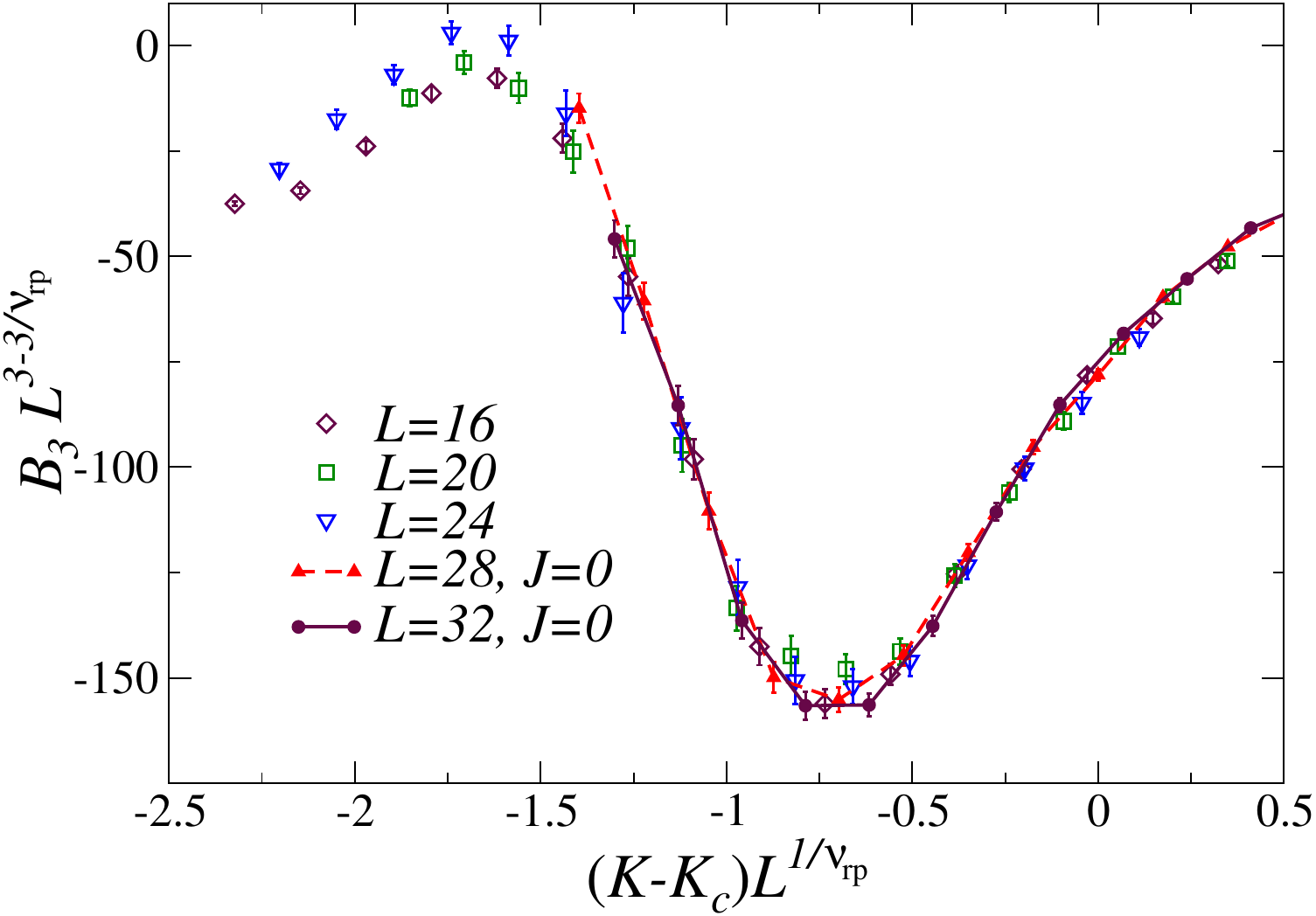}
  \includegraphics[width=0.95\columnwidth]{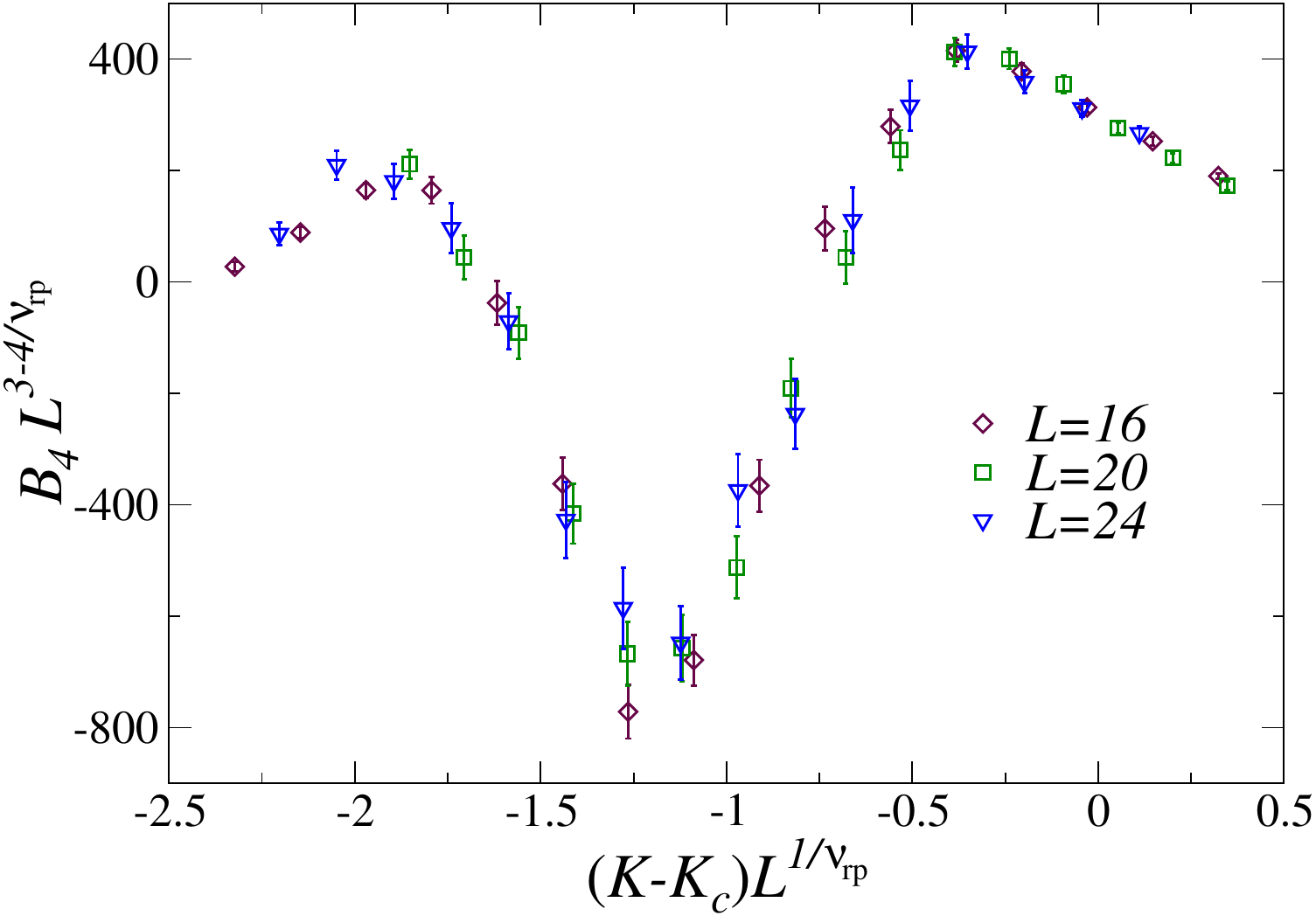}
  \caption{Scaling of the third and of the fourth cumulants for the
    model with $J=0.1$ and $q=0.015$. The critical exponent $\nu_{\rm
      rp}=0.82$ of the RP${\mathbb Z}_2$G universality class has been
    used, together with the optimal estimate $K_c=0.894$ of the
    critical point. For comparison we also report in the upper panel
    data for the third cumulant computed in the RP${\mathbb Z}_2$G
    model obtained in Ref.~\cite{BV-26}, fixing the nonuniversal
    normalizations of the scaling function ${\cal B}_3$ by multiplying
    by 1.0 and 18.8 along the horizontal and vertical directions
    respectively.  }
  \label{fig_J0p1}
\end{figure}

These results for $J=0$ naturally lead us to the hypothesis that the
line of topological transitions starting from $J=0$ persists for
sufficiently small values of $q$, namely $q\lesssim0.03$, becoming of
the RP${\mathbb Z}_2$G universality class.  For larger values of $q$
this low-$J$ transition line should instead disappear.

These general expectations are nicely confirmed by simulations carried
out for $J=0.1$ and $q=0.015$.  Using approximately 200 disorder
samples, and performing for each disorder sample $2\times 10^5$
Metropolis updates of site and bond variables on the whole lattice (we
checked that discarding one quarter of them was sufficient to
guarantee thermalization in all cases), we obtain for the scaling of
the third and fourth cumulants the results reported in
Fig.~\ref{fig_J0p1}. The critical value $K_c$ has been determined by
optimizing the data collapse using Eq.~\eqref{Hg3-scaling} (written as
a function of $K$ instead of $J$), with polynomial approximations of
the scaling functions, and the RP${\mathbb Z}_2$G value of the
correlation length critical exponent $\nu_{\rm rp}=0.82$.  The systematical
errors of this procedure have been estimated by trying different fit
ranges, by varying the value of $\omega$ in the interval $[0.5,1]$,
and by excluding from the fit data obtained on small lattices.  We
obtain $K_c=0.894(2)$ from the data of $B_3$, and $K_c=0.894(3)$ from
those of $B_4$, so we consider
\begin{equation}
K_c=0.894(2)
\end{equation}
as our final estimate.  The quality of the data collapse in
Fig.~\ref{fig_J0p1} is very good, and fully support the presence of a
transition in the RP${\mathbb Z}_2$G universality class.  In
Fig.~\ref{fig_J0p1} we also report, for comparison, data for the third
cumulant in the RP${\mathbb Z}_2$G model obtained in
Ref.~\cite{BV-26}, fixing appropriately the nonuniversal
normalizations entering Eq.~\eqref{Hg3-scaling} (we recall that the
scaling functions ${\cal B}_k(X)$ are universal apart from a
multiplicative factor and a normalization of the argument).  Note that
for $q=0.015$ the transition of the RP${\mathbb Z}_2$G model happens
at~\cite{BV-26} $K_c = 0.8940(8)$, thus in the disordered model the
transition line emerging from $J=0$ is almost vertical, as in the pure
(i.e., without disorder) $\mathbb{Z}_2$ gauge Higgs model.

\begin{figure}
  \includegraphics[width=0.95\columnwidth]{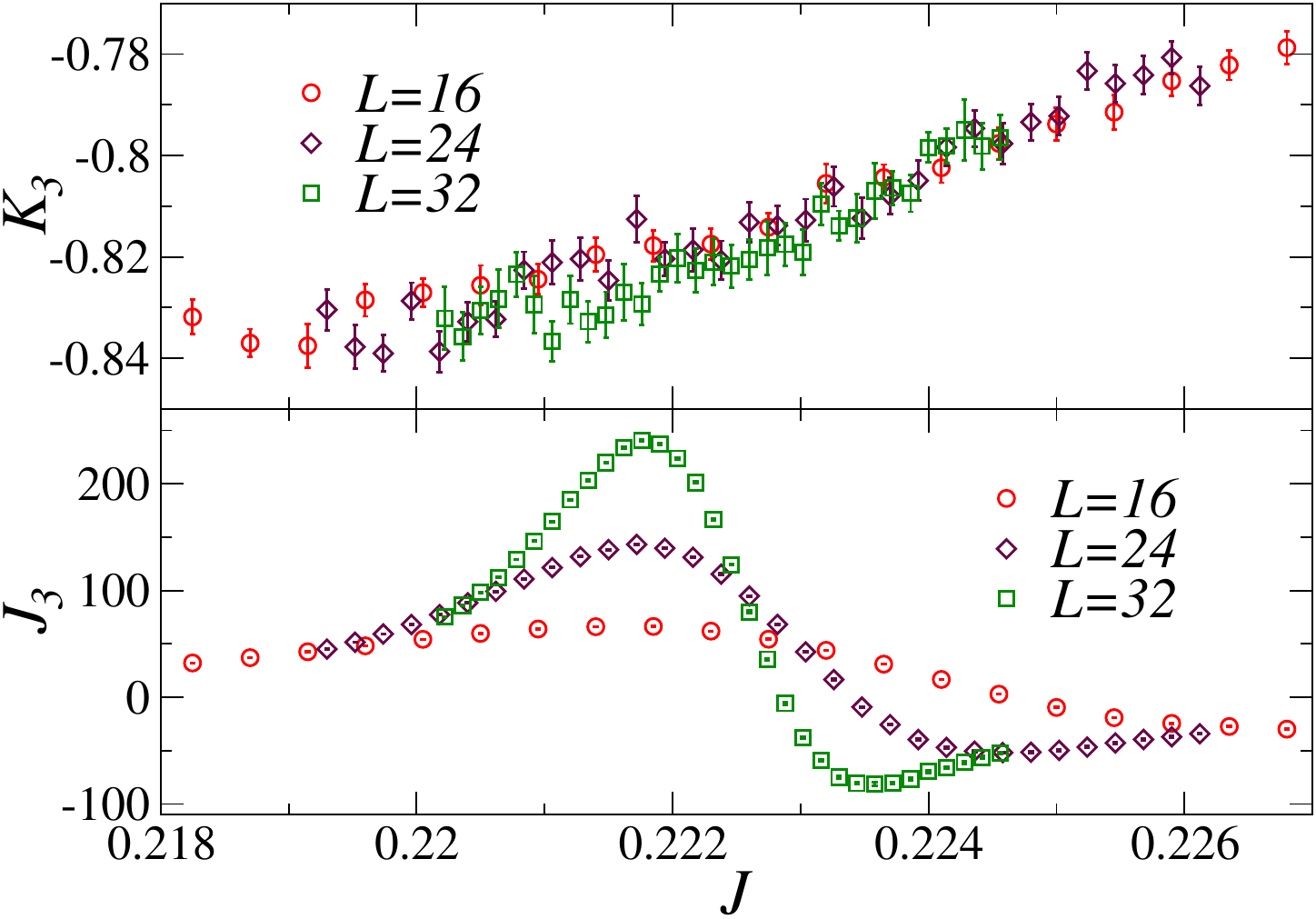}
  \caption{Some results for the lattice ${\mathbb Z}_2$ gauge Higgs model
  without disorder (i.e., $q=0$) and $K=1$, along the Ising$^{\times}$ line.
  Data for the cumulants $K_3=\frac{1}{V}\left(\frac{\partial}{\partial
  K}\right)^3\log Z$ (plaquette variables related) and
  $J_3=\frac{1}{V}\left(\frac{\partial}{\partial J}\right)^3\log Z$ (site
  variables related).}
  \label{fig_K1_q0}
\end{figure}

We now discuss what happens to the Ising$^\times$ transition line of
the pure system, i.e., that ending at a finite $J$ for $K\to\infty$,
when we add quenched random-plaquette disorder. Since the
specific-heat exponent of the 3D Ising universality class (and thus of
the 3D Ising$^{\times}$ universality class) is positive, i.e.,
$\alpha_{\cal I}\approx 0.11$, and the local random quenched disorder
is somehow coupled to an energy term, one could guess this transition
line to change its universality class for $q>0$ also in this case, due
to Harris criterium.  However, along the Ising$^\times$ line the
relevant modes are those associated with the spin variables $s_{\bm
  x}$, bond variables $\sigma_{{\bm x},\mu}$ only playing a secondary
role. Roughly, the only role of the gauge variables is that of
removing the non gauge invariant observables from the theory, without
directly driving the critical behavior (as shown in
Ref.~\cite{BPV-24-unco} the spin correlations along the Ising$^\times$
transition line can be uncovered using an appropriate gauge fixing).

The irrelevance of the plaquettes degrees of freedom at the
Ising$^{\times}$ transition of the pure system can be directly checked
by comparing the results obtained for the cumulants computed by using
derivatives with respect to $K$ (hence cumulants of the plaquette part
of the action), denoted by $K_n$, with those computed by using
derivatives with respect to $J$ (hence cumulants of the spin part of
the action), denoted by $J_n$.  Such a comparison is carried out in
Fig.~\ref{fig_K1_q0} for the third cumulants, from which it clearly
emerges that the plaquette part of the energy is not the most singular
one.  It thus seems reasonable to expect that a quenched disorder only
coupled to the plaquettes does not change the critical behavior,
essentially because the fluctuations of the plaquette contributions
are not the dominant ones along the transition line of the pure
system.

To directly check this hypothesis we performed simulations at $K=1$
for some small values of $q$.  In particular, the FSS analysis of the
energy cumulants for $q=0.01$ and $q=0.015$ turns out to confirm this
simple scenario, showing critical behaviors that are consistent with
those of the standard Ising model, and in particular with those in the
absence of disorder.

\begin{figure}
  \includegraphics[width=0.95\columnwidth]{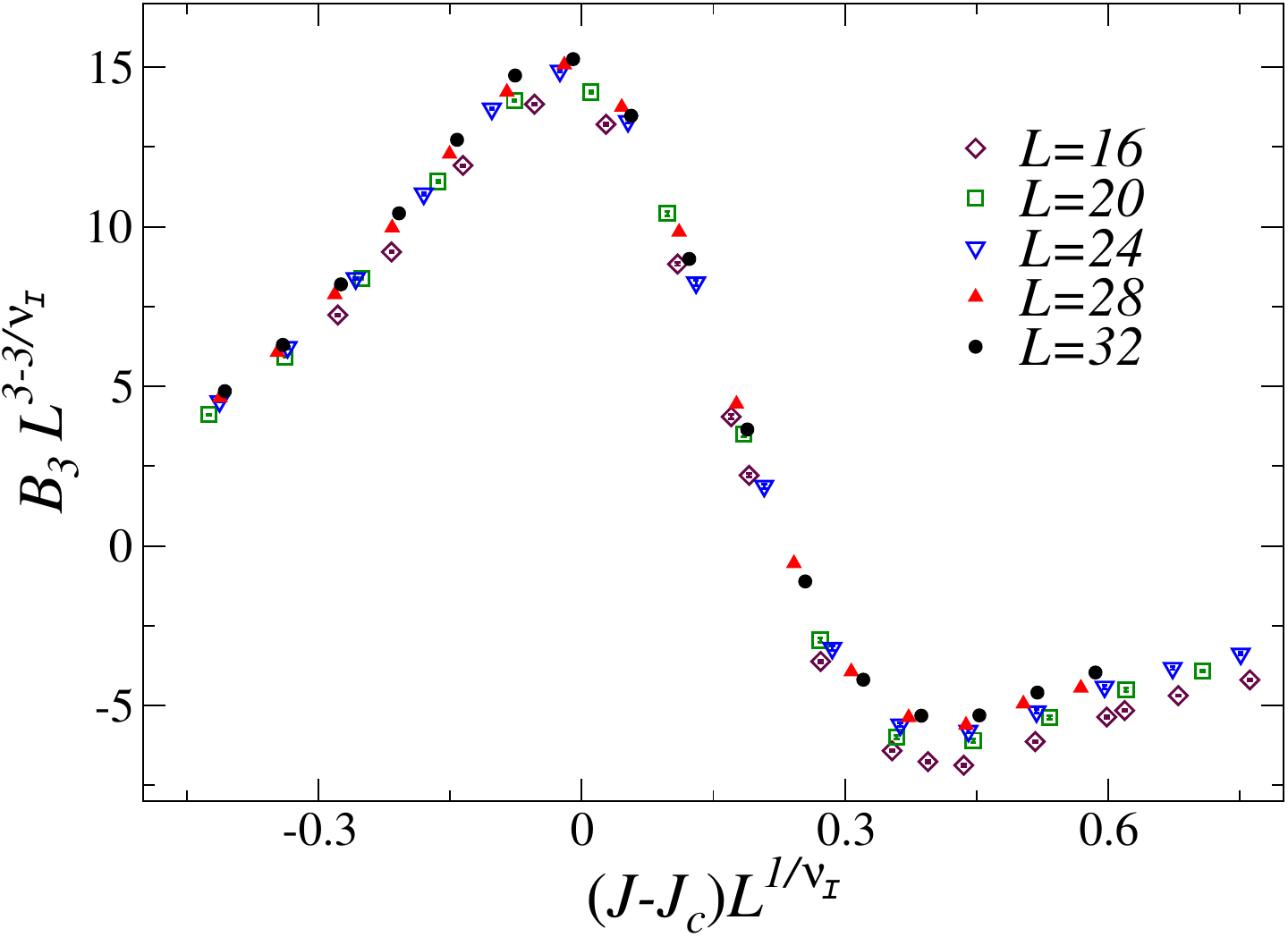}
  \includegraphics[width=0.95\columnwidth]{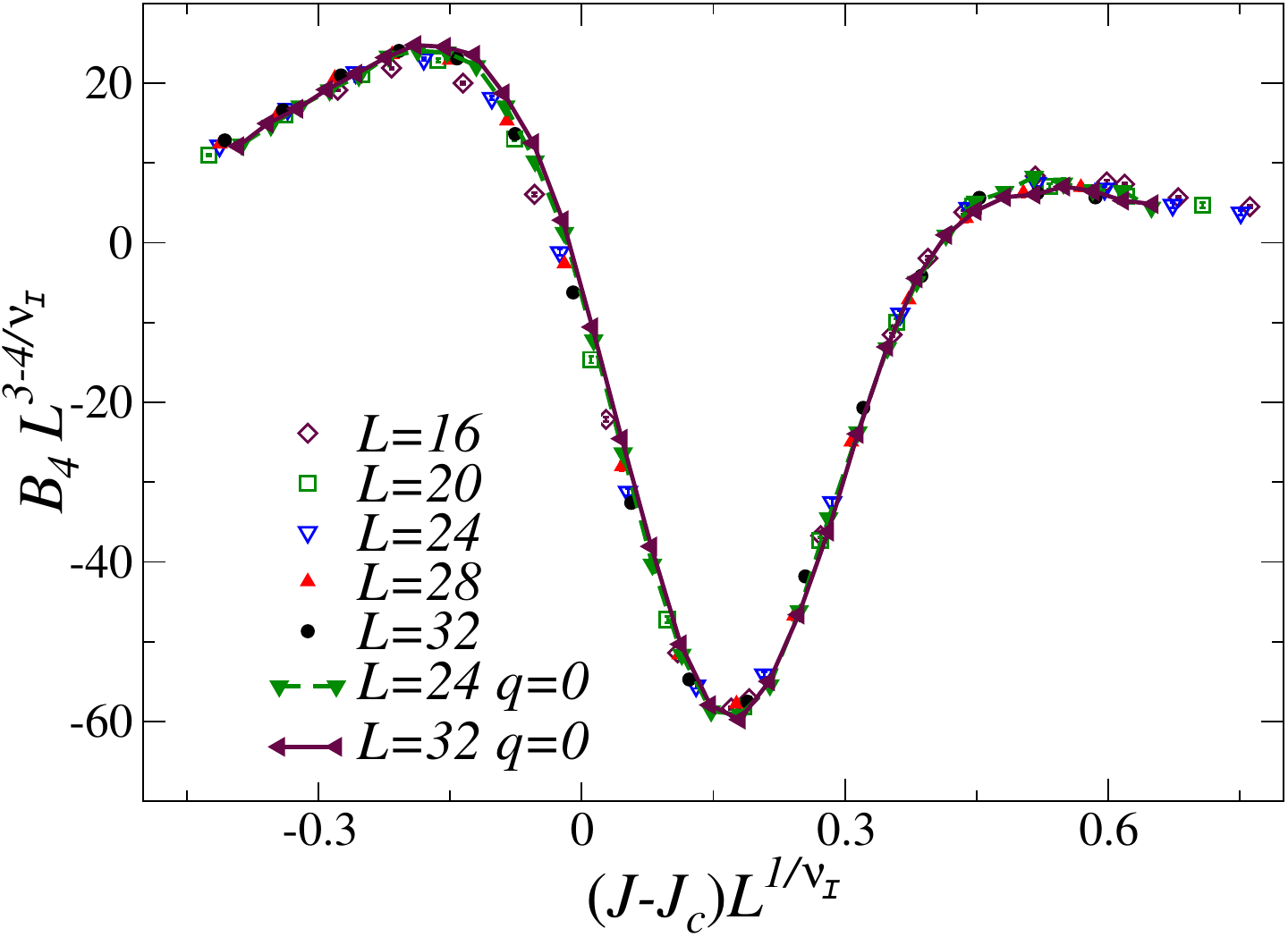}
  \caption{Scaling of the third and of the fourth order cumulants for
    the model with $K=1$ and $q=0.010$. The critical exponent
    $\nu_{\mathcal{I}}=0.629971$ of the Ising universality class has
    been used, together with $J_c=0.22266$.  For comparison we also
    report in the lower panel data for the fourth cumulant computed
    without disorder at $K=1$, rescaled using $\nu_{\cal I}$ and
    $J_c=0.22185$ (see \cite{BPV-24-unco}), multiplied by 0.98 and 87
    along the horizontal and vertical directions, respectively.}
  \label{fig_K1_q0p010}
\end{figure}

In Fig.~\ref{fig_K1_q0p010} we report results for the scaling of the
third and the fourth cumulants obtained for the model with $K=1$ and
$q=0.01$ (the number of samples collected is roughly the same of the
case $J=0.1$ discussed before).  The critical value of the coupling
has been estimated by using the same technique already discussed above
for the model with $J=0.1$.  Using biased fits, with the correlation
length critical exponent $\nu$ and the correction to scaling exponent
$\omega$ fixed to their 3D Ising values, we obtain the results
\begin{equation}
\begin{aligned}
& J_c=0.22263(3)\quad\mathrm{from\ }B_3,\\
& J_c=0.22266(2)\quad\mathrm{from\ }B_4,
\end{aligned} 
\end{equation}
where the first estimate has been obtained using $B_3$ data, the second 
one using $B_4$ data. Using unbiased fits we instead find 
\begin{equation}
\begin{aligned}
& J_c=0.22263(5), & & \nu=0.63(1),  & &\mathrm{from\ }B_3,\\
& J_c=0.22266(2), & & \nu=0.625(4), & &\mathrm{from\ }B_4.\\
\end{aligned}
\end{equation}
The nice scaling observed in Fig.~\ref{fig_K1_q0p010} shows that the
transition is of the Ising$^{\times}$ universality class even for
nonvanishing (small) disorder.  The same conclusion can be reached
also using the estimates of the critical exponent $\nu$ reported
above, with the nice agreement of the different estimates signaling
that systematic errors are well under control.  We explicitly note
that for $B_3$ larger scaling corrections are observed than for $B_4$,
which suggests that the background term in Eq.~\eqref{Hg3-scaling} is
not negligible.

\begin{figure}
  \includegraphics[width=0.95\columnwidth]{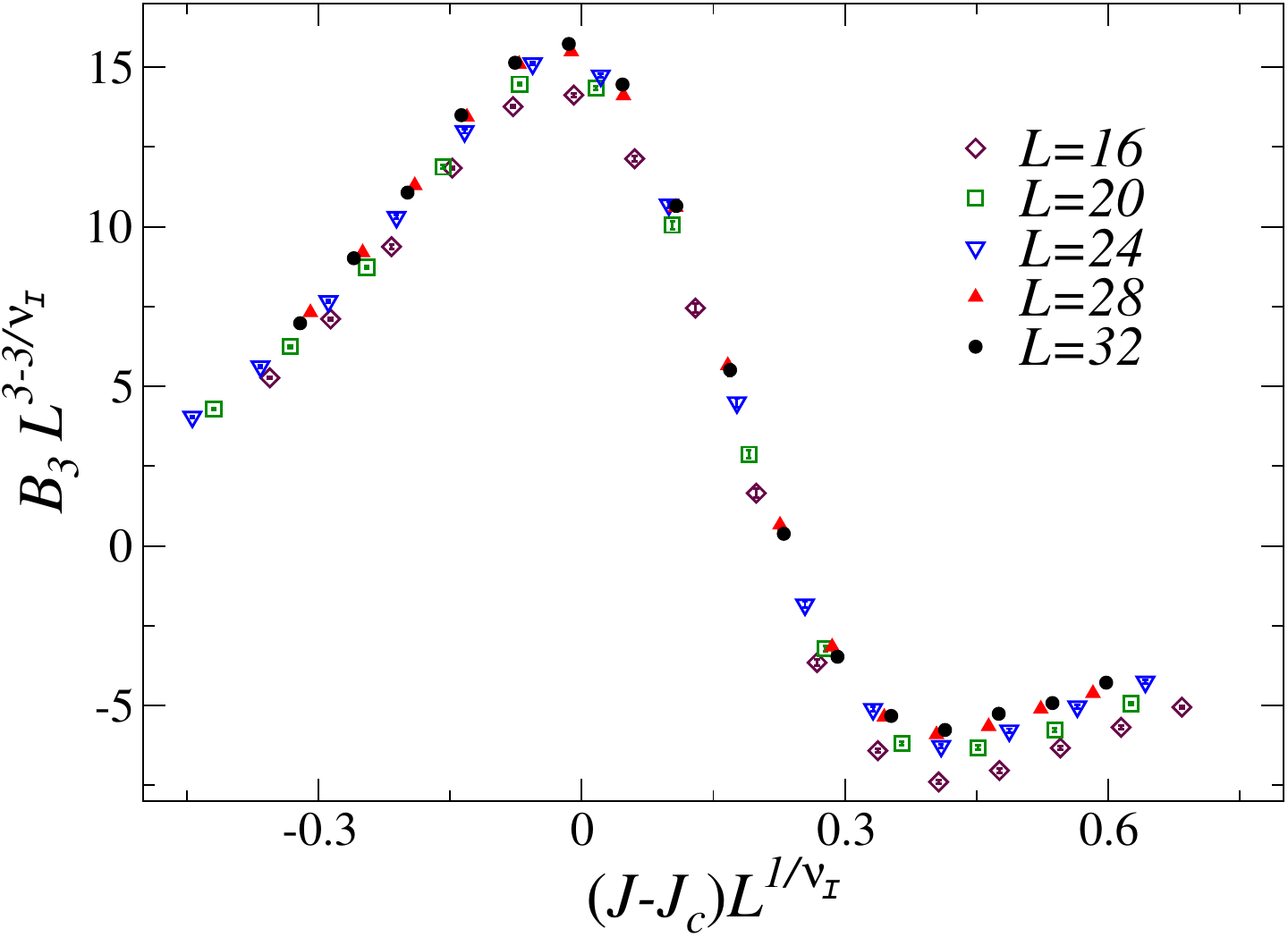}
  \includegraphics[width=0.95\columnwidth]{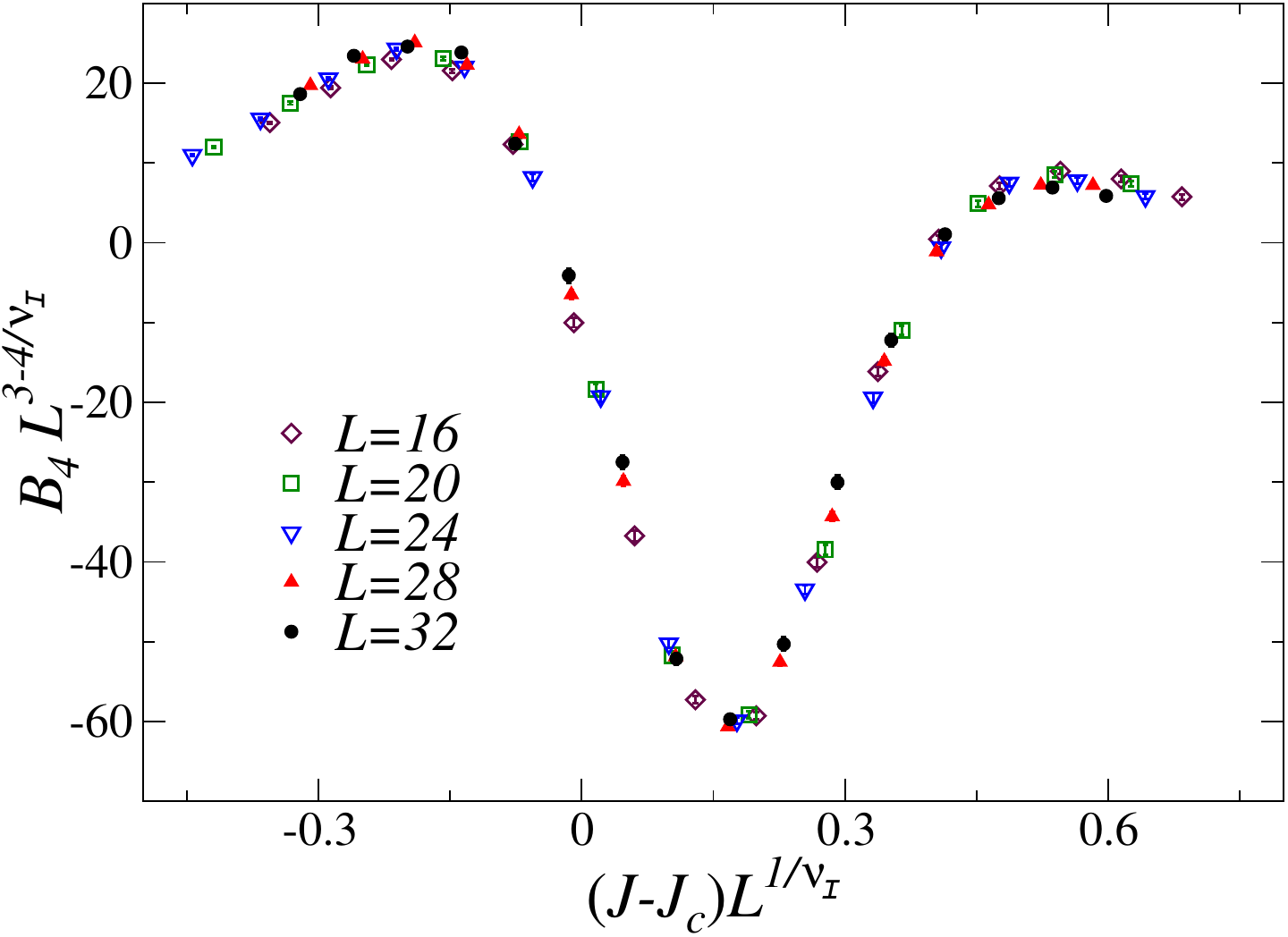}
  \caption{Scaling of the third and of the fourth order cumulants for
    the model with $K=1$ and $q=0.015$. The critical exponent
    $\nu_{\mathcal{I}}=0.629971$ of the Ising universality class has
    been used, together with $J_c=0.22336$.}
  \label{fig_K1_q0p015}
\end{figure}

\begin{figure}
  \includegraphics[width=0.95\columnwidth]{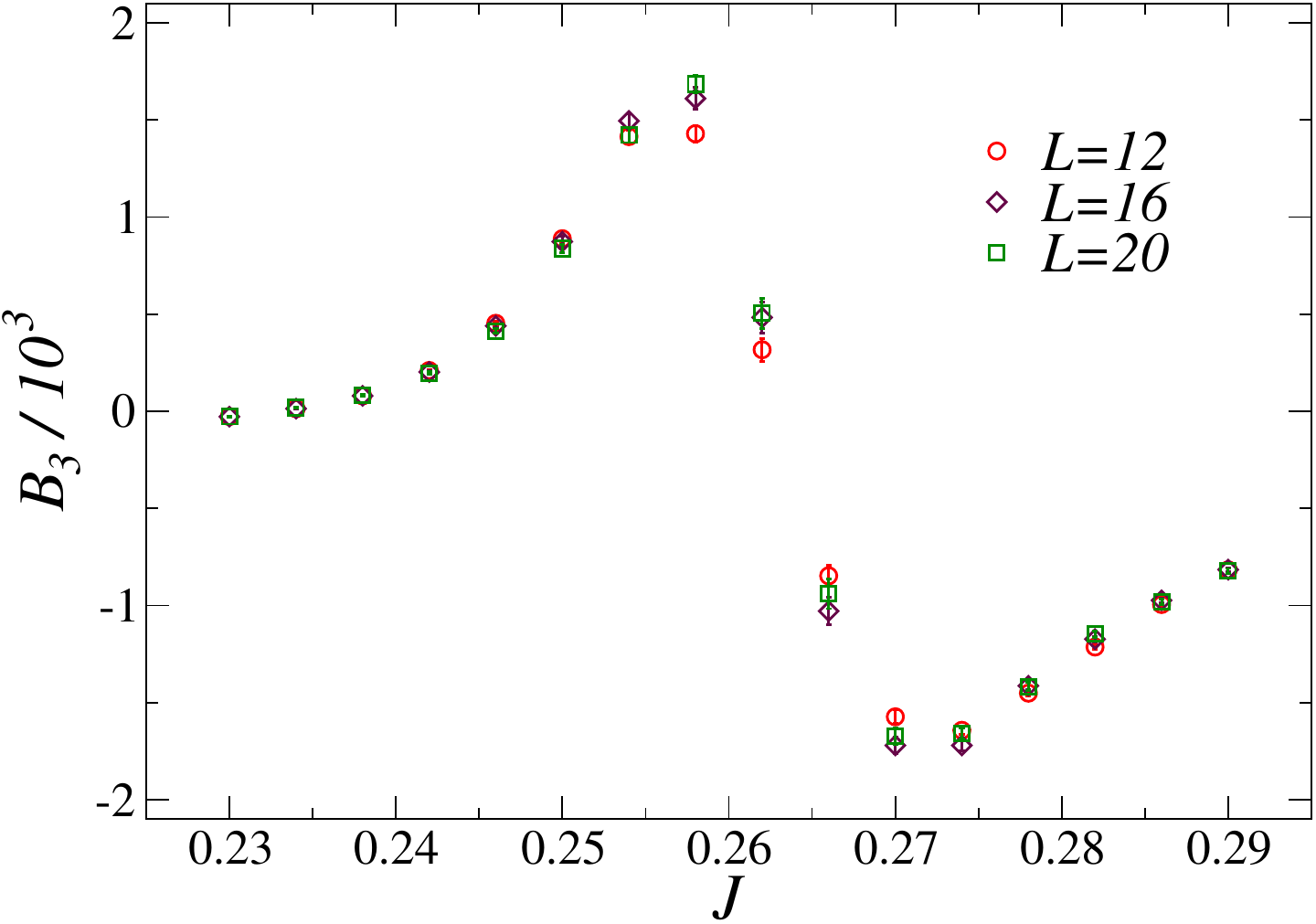}\\
  \vspace{0.2cm}
  \includegraphics[width=0.95\columnwidth]{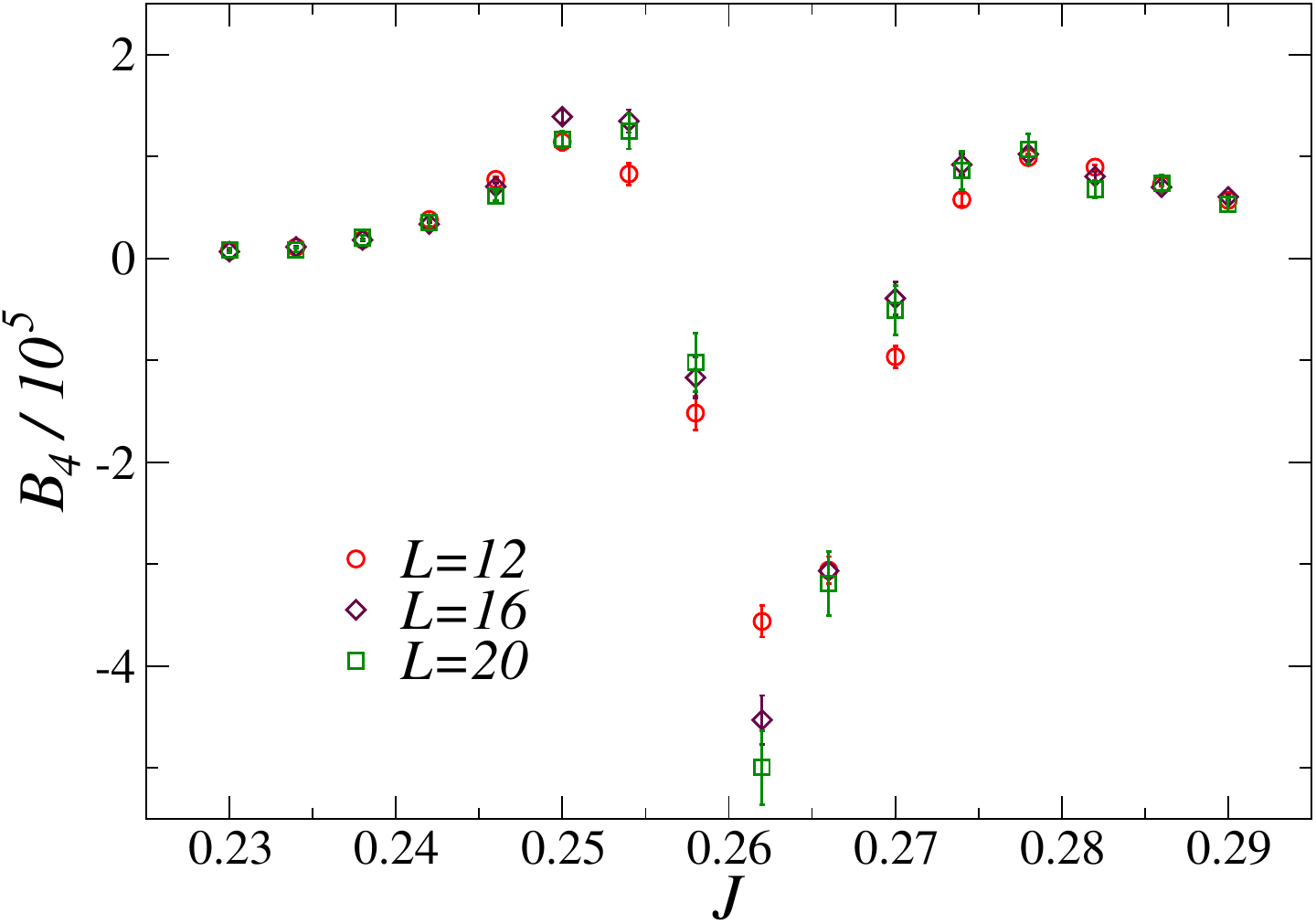}
  \caption{Third and the fourth order cumulants data for the 
  model with $K=1$ and $q=0.05$.}
  \label{fig_K1_q0p05}
\end{figure}

As a final check we verified that the $B_4$ scaling curve is
consistent, up to nonuniversal multiplicative factors, with that of
the pure model (i.e., $q=0$) at $K=1$: in Fig.~\ref{fig_K1_q0p010} we
report data obtained using lattices $L=24$ and $L=32$ for the $q=0$,
$K=1$ model, rescaled using the critical exponent $\nu_{\cal I}$ of
the 3D Ising model and $J_c=0.22185(10)$ (see
Ref.~\cite{BPV-24-unco}), and multiplied by 0.98 and 87 along the
horizontal and vertical directions, respectively (we show such
  a comparison only for $B_4$, because $B_3$ presents much larger
  scaling corrections). The agreement between data at $q=0$ and at
  $q=0.01$ is good, with small discrepancies that can reasonably be
  attributed to the presence of scaling corrections.

A similar investigation has been carried out also for $K=1$ and
$q=0.015$. The outcomes of the biased analyses (i.e., obtained by
fixing $\nu=\nu_{\cal I}$ and $\omega=\omega_{\cal I}$) for the
critical coupling are
\begin{equation}
\begin{aligned}
&J_c=0.22336(6)\quad \mathrm{from\ }B_3,\\ 
&J_c=0.22338(7)\quad \mathrm{from\ }B_4.
\end{aligned} 
\end{equation}
Unbiased estimates give instead 
\begin{equation}
\begin{aligned}
&J_c=0.22335(5), & &\nu=0.63(1),& &\mathrm{from\ }B_3,  \\
&J_c=0.22337(5), & &\nu=0.63(1),& &\mathrm{from\ }B_4,
\end{aligned}
\end{equation}
and the scaling curves for the third and fourth cumulants are shown in
Fig.~\ref{fig_K1_q0p015}. Therefore, also in this case, we 
confirm that the plaquette disorder does not destabilize the
Ising$^{\times}$ universality class.

We have also performed MC simulations at $K=1$ for larger values of
$q$, in particular $q=0.05$. In this case no evidence of transitions
are found, see Fig.~\ref{fig_K1_q0p05}, which suggests that for
sufficiently large values of $q$ also this transition line disappears,
as the topological one starting from the $J=0$ line.

\section{Phase diagram in the presence of randomly-dilute disorder}
\label{rdidis}

We now discuss the second type of disorder, cf. Eq.~(\ref{HiggsHrd}),
modeling the random presence of site defects.  For this purpose it is
useful to recall that the 3D randomly-dilute Ising (RDI) model is
described by the Hamiltonian
\begin{equation}
    H_{\rm rdI} = - J \sum_{{\bm x},\mu} \rho_{\bm x} \, \rho_{{\bm
        x}+\hat{\mu}} \, s_{\bm x} \, s_{{\bm x}+\hat{\mu}},
\label{HiggsHrdi}
\end{equation}
with the probability distribution (\ref{probdisite}) for the quenched
variables $\rho_{\bm x}$, controlled by the defect concentration
$q_s$.  According to the Harris
criterium~\cite{Harris-74,Ma-book,Cardy-book,Vojta-19}, the disorder
associated with the random dilution makes the Ising critical behavior
of the pure system ($q_s=0$) unstable. The critical behaviors for
$q_s>0$ belong to the 3D RDI universality class, whose critical
exponents are given by $\nu_{\rm rdi}=0.683(2)$, see, e.g.,
Refs.~\cite{HPPV-07-rdi,PV-00,PV-02}. Note that such transitions occur
for sufficiently small values of $q_s$, below the percolation point
$q_s^*$ of the simple cubic lattice, therefore for~\cite{LZ-98}
\begin{equation}
0<q<q_s^*=0.6883919(13),
\label{qqs}
\end{equation}
while the ferromagnetic phase transition disappears for
$q_s>q_s^*$. Moreover, the critical value of the coupling $J$ entering
Eq.~\eqref{HiggsHrdi} diverges as $q\to q_s^{*}$, see
Ref.~\cite{BFMMPR-98}.

The behavior of the RDI models leads to the naturally hypothesis that
the Ising$^\times$ transition line of the pure system ($q_s=0$) turns
into a RDI$^\times$ line, where the critical behaviors belong to the
RDI universality class, thus characterized by the critical exponent
$\nu_{\rm rdi}=0.682(3)$~\cite{HPPV-07-rdi}. Note that also the
dynamic exponent $z$ related to the purely relaxational dynamics is
expected to change accordingly, becoming that of the RDI universality
class, which is given by~\cite{HPV-07-zrdi} $z=2.35(2)$.

On the other hand, we note that the randomly-dilution disorder enters only the
term containing the site spins $s_{\bm x}$, therefore it does not affect the
behavior at $J=0$, which remains that of the pure lattice ${\mathbb Z}_2$ gauge
model, characterized by the Ising exponent $\nu_{\cal I}\approx 0.630$. Like
the pure $q_s=0$ case, we do not expect that the critical behavior can change
for finite (sufficiently small) values of $J$.  Therefore, the presence of
random dilution is not expected to change the critical behaviors along the
topological transition line, which remains characterized by the critical
exponent $\nu_{\cal I}$ analogously to the case for $J=0$. Actually, with
increasing $q_s$, this topological transition line starting from $[K_{{\mathbb
Z}_2g},J=0]$ should extend to larger and larger values of $J$, in particular
for $q_s> q_s^*$, where the random dilution should make the effects of the
first term of the Hamiltonian (\ref{HiggsHrd}) negligible for any $J$.  Finally
we mention that also the relaxational critical dynamics is expected to remain
characterized by the same exponent $z=2.610(15)$ of the pure ${\mathbb Z}_2$
gauge model~\cite{BPV-25-dyn,BPV-25-rel,XCMCS-18,BKKLS-90}.

\section{Conclusions}
\label{conclu}

We have reported a study of the effects of uncorrelated quenched
disorder to the phase diagram and continuous transitions of 3D lattice
${\mathbb Z}_2$ gauge Higgs models. We consider two types of quenched
disorder, associated with the sites and plaquettes of the cubic
lattice.  We show that the structure of the phase diagram stays
unchanged for sufficiently small disorder, instead the universality
classes of some of the continuous transitions change. Since both the
continuous transitions lines of the pure model have positive
specific-heat critical exponent $\alpha$ ($\alpha_{\cal I}\approx
0.11$ for 3D Ising-like transitions), using Harris criterium one could
have guessed that both lines change their universality class when
disorder is present.  However only the topological $\mathbb{Z}_2$
transition line emerging from $J=0$ changes its universality class
when plaquette disorder is present, and only the Ising$^{\times}$
transition line departing from $K=\infty$ changes its universality
class when site disorder is present.  This behavior can be explained
by noting that the plaquette coupling is irrelevant at the
Ising$^{\times}$ transition of the pure system, and by duality the
link coupling is irrelevant at the topological $\mathbb{Z}_2$
transition.  Such a scenario has been directly tested by numerical FSS
analyses of MC simulations for the plaquette disorder case, whose
results confirm that the universality class of the transition changes
from $\mathbb{Z}_2$ gauge to RP$\mathbb{Z}_2$G, while the
Ising$^{\times}$ transition is robust against weak plaquette disorder.

For sufficiently weak disorder, i.e., when both continuous transition
lines survive, they are expected to meet at some point, where a
multicritical behavior may develop, like the case of the pure
system. However, to understand the effect of disorder at such MCP is a
much harder numerical task. In the presence of disorder no duality
relation exists, indeed the two transition lines are associated with
different universality class are present. Therefore, we do not have
sufficiently precise information on the location of such intersection
point, unlike the pure case where we know that the MCP must lie on the
self-dual line, making numerical investigations extremely
challenging. Also the fate of the discontinuous transitions line in
the pure system, see Fig.~\ref{phadia}, is unclear.  These questions
call for further investigations.

What happens in the limit of strong disorder is another subject that
deserves to be more carefully investigated. Let us consider for
example the case of site disorder. As the disorder concentration
approaches the percolation threshold the critical value
$J_c(K=\infty)$ diverges, so two different cases are possible: either
the transition lines disappear (and a single phase is thus present) or
the transition line departing from $K=0$ has to reach the top-right
corner of the phase diagram. Note indeed that, even in the presence of
disorder, the model is trivial along the $K=0$ and $J=\infty$ lines,
and no transition can take place there. A completely analogous
reasoning can be done for plaquette disorder, in which case $K_c(J=0)$
diverges for $q\approx 0.03$: also in this case either a single
thermodynamic phase exists or a peculiar reentrant phase should be
present in the bottom-right corner of the phase diagram.

Finally, we mention that one may also consider the case in which both
the classes of disorder are simultaneously present. We have not
presented numerical results for this model. However, on the basis of
the results obtained for the distinct cases, we expect that the phase
diagram show two phases for sufficiently weak disorder, and in
particular a topologically ordered phase, but both continuous
transition lines change their universality class, becoming
RP$\mathbb{Z}_2$G (the one departing from $J=0$) and RDI$^\times$ (the
one departing from $K=\infty$) transition lines.

\acknowledgments

Numerical simulations have been performed on the CSN4 cluster of the Scientific
Computing Center at INFN-PISA.

\end{document}